\documentclass[11pt,letterpaper]{article}
\usepackage{jheppub}
\usepackage[T1]{fontenc}
\usepackage[latin9]{inputenc}
\usepackage{xcolor}
\usepackage{amsmath}
\usepackage{amssymb}
\usepackage{graphicx}

\makeatletter

\pdfoutput=1
\usepackage[colorlinks=true,citecolor=blue,linkcolor=blue]{hyperref}
\usepackage{slashed}

\allowdisplaybreaks

\makeatother

\usepackage{babel}
\begin{document}
	
\title{Probing the $\nu_{R}$-philic $Z'$ at DUNE near detectors}



	
\author[a,b]{Garv Chauhan,} 
\author[b]{P. S. Bhupal Dev,} 
\author[c]{Xun-Jie Xu}
\affiliation[a]{Centre for Cosmology, Particle Physics and Phenomenology (CP3), 
	Universit\'{e} catholique de Louvain, Chemin du Cyclotron 2, 
	B-1348 Louvain-la-Neuve, Belgium}
\affiliation[b]{Department of Physics and McDonnell Center for the Space Sciences,  Washington University, \\ 
	St.\ Louis, MO 63130, USA} 
\affiliation[c]{Institute of High Energy Physics, Chinese Academy of Sciences, Beijing 100049, China}
\emailAdd{garv.chauhan@uclouvain.be,  bdev@wustl.edu, xuxj@ihep.ac.cn}

\abstract{\noindent We consider a hidden $U(1)$ gauge
symmetry under which only the right-handed neutrinos ($\nu_{R}$) are charged. The corresponding gauge boson is referred to as the $\nu_{R}$-philic $Z'$.   
Despite the absence of direct gauge couplings to ordinary matter
at tree level, loop-induced couplings of the $\nu_{R}$-philic $Z'$
via left-right neutrino mixing can be responsible for its experimental
accessibility. An important feature of the $\nu_{R}$-philic $Z'$
is that its couplings to neutrinos are generally much larger than its couplings
to charged leptons and quarks, thus providing a particularly interesting
scenario for future neutrino experiments such as
DUNE to probe. We consider two approaches to probe the $\nu_{R}$-philic
$Z'$ at DUNE near detectors via (i) searching for $Z'$ decay signals,
and (ii) precision measurement of elastic neutrino-electron scattering mediated by the $Z'$ boson.
We show that the former will have sensitivity comparable to or better
than previous beam dump experiments, while the latter will improve current limits substantially for large neutrino couplings.}


\maketitle

\section{Introduction}
\noindent 
The discovery of neutrino oscillations has established non-zero neutrino masses on a firm footing~\cite{ParticleDataGroup:2020ssz}. It also clearly points towards the presence of Beyond the Standard Model (BSM) physics. There has been monumental progress to understand the standard $3 \times 3$ neutrino mixing paradigm by measuring the mixing parameters with ever-increasing accuracy but we are yet to understand the neutrino mass mechanism. There have been several proposals in the literature (for reviews, see e.g.~Refs.~\cite{Mohapatra:2006gs, deGouvea:2016qpx}), which always come with new interactions of neutrinos. Some of these ideas might be testable with the upcoming high-intensity frontier experiments in the near future~\cite{DUNE:2020fgq}.

In this work, we consider a BSM scenario with a hidden $U(1)$ gauge symmetry under which only the right-handed neutrinos ($\nu_R$) are charged~\cite{Chauhan:2020mgv}. These right-handed neutrinos interact with SM particles only through their mixing with the active neutrinos. As a result, the new gauge boson $Z'$ associated with the hidden $U(1)$ symmetry can interact with the SM particles only through the left-right neutrino mixing and through $W^{\pm}/Z$-loop induced couplings (see Fig.~\ref{fig:loop}). A noteworthy feature of the $\nu_{R}$-philic $Z'$ is that its couplings to neutrinos (including the light ones via mixing) are generically larger than its loop-induced couplings to electrons. This is directly useful for probing this scenario at the intensity frontier neutrino experiments like DUNE~\cite{DUNE:2020ypp}. Therefore, unlike most $Z'$ models for which there are already severe constraints from beam dump experiments and $e^{+}e^{-}$ colliders, and hence, little space for DUNE to probe, the $\nu_{R}$-philic $Z'$ presents significantly better prospect in neutrino experiments. 
In our case, due to stronger couplings to neutrinos than to quarks, $Z'$ is primarily produced through charged pion decays instead of neutral pion decays (see also a similar situation in Ref.~\cite{Bakhti:2018avv} for the production of neutrinophilic $Z'$). This is in stark contrast to the typical $Z'$ models in literature, e.g. $B-L$ model, where neutral pion decay is the dominant production mode for light $Z'$~\cite{Berryman:2019dme, Dev:2021qjj}. 

In this work, we study the detection prospects for the $\nu_R$-phillic $Z'$ model at the near detector complex of DUNE~\cite{DUNE:2021tad} in two different ways: (i) elastic $\nu$-$e$ scattering, and (ii) using DUNE as a beam dump experiment. 
The process of elastic $\nu$-$e$ scattering has been widely used to constrain new interactions of neutrinos in the 
literature~\cite{Bilmis:2015lja,Lindner:2018kjo,Bischer:2018zcz,Ballett:2019xoj,Link:2019pbm,Dev:2021xzd, Chakraborty:2021apc}, due to its rather small theoretical uncertainties compared to the neutrino-nucleus scattering, 
although the neutrino-electron scattering cross section is  much smaller than the neutrino-nucleus scattering cross section.\footnote{Since the successful observation of coherent elastic neutrino-nucleus scattering (CE$\nu$NS)~\cite{COHERENT:2017ipa}, neutrino-nucleus scattering has also been used to constrain new physics, see e.g.~\cite{Lindner:2016wff,Dent:2017mpr,Farzan:2018gtr,Abdullah:2018ykz,Brdar:2018qqj,Cadeddu:2018dux}.} 
Nevertheless, a large neutrino flux at DUNE offers a  significant statistical advantage to probe BSM physics at the near detector. 
In addition to neutrino scattering, the $Z'$ can be directly produced from proton bremsstrahlung and decay of charged/neutral mesons produced from proton beam striking the target. In this way, DUNE can also be modeled as a beam dump experiment~\cite{Berryman:2019dme}.

A number of studies have shown that beam dump experiments in general have great capabilities to constrain $Z'$ with very weak couplings, provided that its visible decay width is not too small~\cite{Bjorken:2009mm,Andreas:2012mt,Ilten:2018crw,Bauer:2018onh,Coy:2021wfs}. For the $\nu_R$-philic $Z'$, the comparatively large couplings to neutrinos would suppress the visible decay width but enhance the production rate. Both aspects will be taken into account in this work.

The rest of paper is organized as follows: In Sec.~\ref{sec:basic}, we discuss the framework for the neutrinophilic scenario considered in this work. In Sec.~\ref{sec:DUNEnue}, we study prospects for this scenario at DUNE through neutrino-electron scattering.  In Sec.~\ref{sec:DUNEbd}, we analyze the role of DUNE as a beam dump experiment in constraining the neutrinophilic scenario.  In Sec.~\ref{sec:Results}, we discuss the implications of the combined analysis from neutrino scattering and beam dump. Finally we conclude in Sec.~\ref{sec:Conclusion}. Some details on the $Z'$ production via proton bremsstahlung are relegated to Appendix~\ref{sec:proton-brem}. 






\section{Model Framework \label{sec:basic}}


We consider a $Z'$ vector boson that is coupled to $n$ right-handed
neutrinos ($\nu_{R}$) as follows:\footnote{Following the same notations as in Ref.~\cite{Chauhan:2020mgv}, we
	adopt two-component Weyl spinors for all chiral fermions.}
\begin{equation}
{\cal L}\supset g'Z_{\mu}'\sum_{i=1}^{n}\nu_{R,i}^{\dagger}\overline{\sigma}^{\mu}Q_{R,i}\nu_{R,i}\thinspace,\label{eq:-1}
\end{equation}
where $g'$ is the gauge coupling and $Q_{R,i}$ is the $U(1)'$
charge of $\nu_{R,i}$. For brevity, we will rewrite Eq.~\eqref{eq:-1} in the matrix form:
\begin{equation}
{\cal L}\supset g'Z_{\mu}'\nu_{R}^{\dagger}\overline{\sigma}^{\mu}Q_{R}\nu_{R}\thinspace,\label{eq:-28}
\end{equation}
with $Q_{R}={\rm diag}(Q_{R,1},\ Q_{R,2},\ Q_{R,3},\cdots)$ and $\nu_{R}=(\nu_{R,1},\ \nu_{R,2},\ \nu_{R,3},\cdots)^{T}$.

If $\nu_{R}$ are the only fermions charged under $U(1)'$, the cancellation
of chiral anomalies
requires $\sum_{i}Q_{R,i}^{3}=0$.\footnote{
	Unlike other $U(1)$ extensions of the SM,  mixed gauge anomalies are absent here because no fermions are charged under both new and SM gauge symmetries. Hence we only need to consider triangle diagrams with purely $U(1)'$ gauge boson legs.
}  
Therefore, 
one often introduces pairs of $\nu_{R}$ with opposite charges to
attain the cancellation~\cite{Chang:2011kv,Ma:2013yga,Lindner:2013awa,Farzan:2016wym,Abdallah:2021npg,Abdallah:2021dul}. Alternatively, one may consider other
charge assignment such as $Q_{R}=(1,\ 1,\ 1,\ -4,\ -4,\ 5)$ or $(3,\ 4,\ 5,\ -6)$
which also satisfies $\sum_{i}Q_{R,i}^{3}=0$. 
We refrain from further
discussions on the charge assignment as this can be rather model-dependent
while the results in this work can be simply rescaled by a factor
of ${\cal O}(Q_{R})$.

The $\nu_{R}$ sector is connected to left-handed neutrinos $\nu_{L}$
via Dirac mass terms. In addition, $\nu_{R}$ may have Majorana mass
terms. Formulated in the matrix form,  the Dirac and Majorana mass
terms are given by
\begin{equation}
{\cal L}\supset\nu_{L}^{T}m_{D}\nu_{R}+\frac{1}{2}\nu_{R}^{T}m_{R}\nu_{R}+{\rm h.c.}\thinspace,\label{eq:-2}
\end{equation}
where $m_{D}$ is a $3\times n$ matrix and $m_{R}$ is a $n\times n$
matrix.  The Dirac mass term has to break the $U(1)'$ symmetry, which can be achieved by adding another Higgs doublet charged under $U(1)'$, as discussed in Ref.~\cite{Chauhan:2020mgv}.  Alternatively, introducing new scalar singlets charged under $U(1)'$ and heavy neutral fermions free of all charges can effectively produce such terms at low energies. Our analyses in this work based on Eq.~\eqref{eq:-2} would not be affected by these model-dependent details.

\subsection{Loop-induced coupling\label{subsec:Loop-induced-coupling}}

\begin{figure}
	\centering
	
	\includegraphics[width=0.7\textwidth]{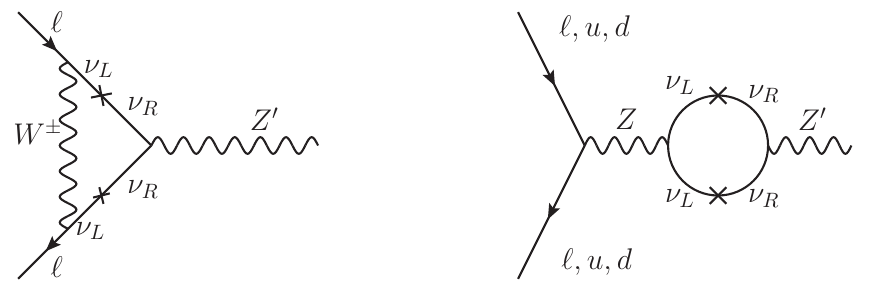}
	
	\caption{Loop-induced couplings of the $\nu_{R}$-philic $Z'$ to charged leptons
		($\ell$) and quarks ($u$ and $d$). The $\times$ denotes the Dirac mass insertion. \label{fig:loop}}
	
\end{figure}

In the presence of the $Z'$ couplings given by Eq.~\eqref{eq:-28}
and the neutrino mass terms in Eq.~\eqref{eq:-2}, there are loop-induced
couplings of $Z'$ to the SM charged leptons ($\ell$) and quarks
($u$ and $d$) generated by  the diagrams in Fig.~\ref{fig:loop}.
The loop-induced couplings are finite (i.e.~not UV divergent) when
both light and heavy neutrino mass eigenstates are included in the
loop integral. 

For illustration, let us first discuss the case with only one $\nu_{L}$
and one $\nu_{R}$ so that $m_{D}$ and $m_{R}$ are $1\times1$ matrices.
In this case,  the loop-induced couplings read~\cite{Chauhan:2020mgv}:
\begin{equation}
{\cal L}_{{\rm loop}}=\sum_{f}g_{f}f^{\dagger}\overline{\sigma}^{\mu}Z_{\mu}'f\thinspace,\label{eq:-14}
\end{equation}
where $f$ denotes SM chiral fermions and $g_{f}$ is determined as follows:
\begin{equation}
g_{f}=\begin{cases}

g_{f}^{(W)}+g_{f}^{(Z)} & f=\ell_{L}\\
g_{f}^{(Z)} & {\rm otherwise}\\
\end{cases}\thinspace,\label{eq:-39}
\end{equation}
with
\begin{align}
g_{f}^{(W)} & =-\frac{\sqrt{2}G_{F}m_{D}^{2}}{8\pi^{2}}g'Q_{R}\thinspace,\label{eq:-15}\\
g_{f}^{(Z)} & =\frac{\sqrt{2}G_{F}m_{D}^{2}}{8\pi^{2}}g'Q_{R}Q_{f}^{(Z)}\thinspace.\label{eq:-16}
\end{align}
Here the superscripts $(W)$ and $(Z)$ indicate contributions from
the $W$ and $Z$ diagrams in Fig.~\ref{fig:loop}, respectively.
The result of the $Z$ diagram depends on the so-called $Z$ charge
in the SM, defined as  $Q_{f}^{(Z)}\equiv I_{3}-Q_{{\rm em}}s_{W}^{2}$
with $I_{3}$ the isospin and $Q_{{\rm em}}$ the electric charge
(e.g. $Q_{\nu_{L}}^{(Z)}=1/2$, $Q_{e_{L}}^{(Z)}=-1/2+s_{W}^{2}$,
$Q_{e_{R}}^{(Z)}=s_{W}^{2}$), and $s_W\equiv \sin\theta_W$, where $\theta_W$ is the weak mixing angle.  

Eqs.~\eqref{eq:-15} and \eqref{eq:-16} are derived under the approximation
that $m_{D}/m_{R}\ll1$. The results are proportional to $m_{D}^{2}$
which can be understood from Fig.~\ref{fig:loop} since each diagram
needs two mass insertions to connect $\nu_{L}$ and $\nu_{R}$ lines. 
If $Z'$ is replaced with a scalar, then the diagrams would be proportional to $m_R^{-1}$, leading to more suppressed loop-induced couplings~\cite{Xu:2020qek}. 

For $3\nu_{L}+n\nu_{R}$, the loop-induced couplings are also finite
and of the same order of magnitude  as Eqs.~\eqref{eq:-15} and \eqref{eq:-16}.
However, with the most general mass matrices, the loop-induced couplings
cannot be written into compact and simple forms---see Eqs.~(3.8)
and (3.9) in Ref.~\cite{Chauhan:2020mgv}. Besides, a large number
of free parameters in the flavor structure make it more difficult
to obtain simple correlations among the loop-induced couplings. For
simplicity, we concentrate on the case of $n=3$  and assume that
$m_{D}$ and $m_{R}$ can be simultaneously diagonalized as follows:\footnote{   That $m_{D}$ and $m_{R}$ can be simultaneously diagonalized
	is a rather common feature in flavor symmetry models---see e.g. Refs.~\cite{Rodejohann:2015hka,Rodejohann:2017lre, Smirnov:2018luj}.} 
\begin{align}
U_{L}^{T}m_{D}U_{R} & =m_{D}^{(d)}\thinspace,\label{eq:-18}\\
U_{R}^{T}m_{R}U_{R} & =m_{R}^{(d)}\thinspace,\label{eq:-17}
\end{align}
where $U_{L}$ and $U_{R}$ are $3\times3$ unitary matrices, $m_{D}^{(d)}$
and $m_{R}^{(d)}$ are diagonal matrices. 

The full neutrino mass matrix can be diagonalized by a $6\times6$ unitary
matrix $U$:
\begin{align}
\left(\begin{array}{c}
\nu_{L}\\
\nu_{R}
\end{array}\right)& =U\left(\begin{array}{c}
\nu_{1,2,3}\\
\nu_{4,5,6}
\end{array}\right),\\ 
U^{T}\left(\begin{array}{cc}
0 & m_{D}\\
m_{D}^{T} & m_{R}
\end{array}\right)U & ={\rm diag}(m_{1},\ \cdots,m_{6})\thinspace.\label{eq:-19}
\end{align}
Eqs.~\eqref{eq:-18} and \eqref{eq:-17} allow us to write $U$ as
\begin{equation}
U=\left(\begin{array}{cc}
U_{L}\\
& U_{R}
\end{array}\right)U'\thinspace,\ \ U'=\left(\begin{array}{cc}
-i\thinspace{\rm diag}(c_{1},c_{2},c_{3}) & {\rm diag}(s_{1},s_{2},s_{3})\\
i\thinspace{\rm diag}(s_{1},s_{2},s_{3}) & {\rm diag}(c_{1},c_{2},c_{3})
\end{array}\right),\label{eq:-21}
\end{equation}
where $(s_{i},\ c_{i})=(\sin\theta_{i},\cos\theta_{i})$ and $\theta_{i}=\arctan\sqrt{m_{i}/m_{i+3}}$.

With the above assumption, the loop-induced couplings previously
computed in Ref.~\cite{Chauhan:2020mgv} can be reformulated into
the following matrix form:
\begin{align}
g_{f}^{(W)} & =-\frac{\sqrt{2}G_{F}\left[m_{D}\hat{Q}_{R}m_{D}^{\dagger}\right]_{\alpha\beta}}{8\pi^{2}}g'\thinspace,\label{eq:-15-1}\\
g_{f}^{(Z)} & =\frac{\sqrt{2}G_{F}\thinspace{\rm tr}\left[m_{D}\hat{Q}_{R}m_{D}^{\dagger}\right]}{8\pi^{2}}g'Q_{f}^{(Z)}\thinspace,\label{eq:-16-1}
\end{align}
where $\hat{Q}_{R}\equiv U_{R}{\rm diag}(Q_{R,1},Q_{R,2},Q_{R,3})U_{R}^{\dagger}$
and  $(\alpha,\ \beta)\in\{e,\ \mu,\ \tau\}$ are flavor indices.
Eq.~\eqref{eq:-15-1} implies that flavor-changing couplings could
be generated by the $W$ diagram. It can be flavor diagonal if the
charge assignments $(Q_{R,1},Q_{R,2},Q_{R,3})$ and $m_{D}^{(d)}$
satisfy 
\begin{equation}
\left[m_{D}^{(d)}\right]^{2}{\rm diag}(Q_{R,1},Q_{R,2},Q_{R,3})\propto I_{{\rm 3\times3}}\thinspace.\label{eq:-40}
\end{equation}
In this case, we define
\begin{align}
\overline{m}_{D}^{2} & \equiv\frac{1}{3}{\rm tr}\left\{ \left[m_{D}^{(d)}\right]^{2}{\rm diag}(Q_{R,1},Q_{R,2},Q_{R,3})\right\} ,\label{eq:-41}\\
\epsilon_{{\rm loop}} & \equiv\frac{\sqrt{2}G_{F}\overline{m}_{D}^{2}}{8\pi^{2}}\thinspace,
\end{align}
and rewrite Eqs.~\eqref{eq:-15-1} and \eqref{eq:-16-1} as
\begin{align}
g_{f}^{(W)} & =-\epsilon_{{\rm loop}}\thinspace g'\thinspace,\label{eq:-15-1-2}\\
g_{f}^{(Z)} & =3\thinspace\epsilon_{{\rm loop}}\thinspace g'Q_{f}^{(Z)}\thinspace.\label{eq:-16-1-2}
\end{align}

\subsection{Neutrino flavor states at low energies\label{subsec:Neutrino-flavor-states}}

Neutrinos in most experiments are produced via charged-current interactions.
Since $\nu_{L}$ are the eigenstates of such interactions, one
might expect that neutrinos at production are simply $\nu_{L}$. However,
due to a small fraction of heavy neutrino components in $\nu_{L}$,
only a quantum superposition of light mass eigenstates ($\nu_{1}$,
$\nu_{2}$, $\nu_{3}$) can be  produced if the heavy mass eigenstates
are heavier than the masses of the particles responsible for neutrino production
(typically $\pi^{\pm}$, $K^{\pm},$ $\mu^{\pm}$, etc.).   This implies
that a neutrino state produced in the laboratory is slightly different
from $\nu_{L}$.   To account for the difference, we define neutrino
flavor states at low energies as follows:
\begin{equation}
\nu_{\alpha}^{{\rm lab}}=\sum_{i=1}^{3}\left(U_{L}\right)_{\alpha i}\nu_{i}\thinspace,\ \alpha\in\{e,\mu,\tau\}\thinspace.\label{eq:-22}
\end{equation}
Further, we  introduce a set of orthogonal states $\nu_{\alpha}^{\perp{\rm lab}}=\sum_{i=4}^{6}\left(U_{R}\right)_{\alpha i}\nu_{i}$
to complete the basis so that $(\nu^{{\rm lab}},\ \nu^{\perp{\rm lab}})$
and $(\nu_{L},\ \nu_{R})$ are connected by a unitary transformation:
\begin{equation}
\left(\begin{array}{c}
\nu_{L}\\
\nu_{R}
\end{array}\right)=\left(\begin{array}{cc}
-iC_{L} & S\\
iS^{\dagger} & C_{R}
\end{array}\right)\left(\begin{array}{c}
\nu^{{\rm lab}}\\
\nu^{\perp{\rm lab}}
\end{array}\right),\label{eq:-23}
\end{equation}
where
\begin{align}
C_{L,R} & =U_{L,R}{\rm diag}(c_{1},c_{2},c_{3})U_{L,R}^{\dagger}\thinspace,\label{eq:-29}\\
S & =U_{L}{\rm diag}(s_{1},s_{2},s_{3})U_{R}^{\dagger}\thinspace.\label{eq:-30}
\end{align}
Since $c_{i}\approx1$ and $s_{i}\sim m_{D}/m_{R}\ll1$, the unitary
matrix in Eq.~\eqref{eq:-23} performs only a small rotation from
$(\nu^{{\rm lab}},\ \nu^{\perp{\rm lab}})$ to $(\nu_{L},\ \nu_{R})$.
In the basis of $(\nu_{L},\ \nu_{R})$, the $Z'$ couplings to the
neutrino sector (including loop-induced couplings to $\nu_{L}$) read:
\begin{equation}
{\cal L}\supset g'\left(\nu_{L},\ \nu_{R}\right)\left(\begin{array}{cc}
\frac{3}{2}\epsilon_{{\rm loop}}I_{{\rm 3\times3}}\\
& Q_{R}
\end{array}\right)\left(\begin{array}{c}
\nu_{L}\\
\nu_{R}
\end{array}\right)\thinspace.\label{eq:-32}
\end{equation}
Applying the basis transformation in Eq.~\eqref{eq:-23}, we obtain
the effective couplings of $Z'$ to $\nu^{{\rm lab}}$:
\begin{equation}
{\cal L}_{Z',\nu^{{\rm lab}}}=g'Z_{\mu}'\left(\nu^{{\rm lab}}\right)^{\dagger}\overline{\sigma}^{\mu}\left[\frac{3}{2}\epsilon_{{\rm loop}}C_{L}C_{L}^{\dagger}+SQ_{R}S^{\dagger}\right]\nu^{{\rm lab}}\thinspace.\label{eq:-31}
\end{equation}

\subsection{Effective $Z'$ couplings}

Let us summarize the effective couplings of $Z'$ to normal matter
and neutrinos:
\begin{align}
{\cal L} & \supset\overline{\psi_{q}}\left[g_{qL}\gamma^{\mu}P_{L}+g_{qR}\gamma^{\mu}P_{R}\right]Z'_{\mu}\psi_{q}\nonumber \\
& +\overline{\psi_{e}}\left[g_{eL}\gamma^{\mu}P_{L}+g_{eR}\gamma^{\mu}P_{R}\right]Z'_{\mu}\psi_{e}+\overline{\psi_{\nu}}\left[g_{\nu}\gamma^{\mu}P_{L}\right]Z'_{\mu}\psi_{\nu}\thinspace,\label{eq:-42}
\end{align}
where $q=u$ or $d$ and we use $\psi$  to denote four-component
Dirac spinors. For neutrinos, the left-handed part of $\psi_{\nu}$
is the neutrino state produced in the laboratory, i.e.~$P_{L}\psi_{\nu}=\nu^{{\rm lab}}$
where $\nu^{{\rm lab}}$ has been introduced in Sec.~\ref{subsec:Neutrino-flavor-states}.
Combining the previous results, we have the following couplings:
\begin{align}
g_{\nu} & =g'\left[\frac{3}{2}\epsilon_{{\rm loop}}C_{L}C_{L}^{\dagger}+SQ_{R}S^{\dagger}\right],\label{eq:-53-1}\\
\left(g_{e_{L}},\ g_{e_{R}}\right) & =3\epsilon_{{\rm loop}}g'\left(s_{W}^{2}-5/6,\ s_{W}^{2}\right),\label{eq:-53-2}\\
\left(g_{u_{L}},\ g_{u_{R}}\right) & =\epsilon_{{\rm loop}}g'\left(1/2-2s_{W}^{2},\ -2s_{W}^{2}\right),\label{eq:-53-3}\\
\left(g_{d_{L}},\ g_{d_{R}}\right) & =\epsilon_{{\rm loop}}g'\left( s_{W}^{2}-5/2,\ s_{W}^{2} \right).\label{eq:-53-4}
\end{align}
For later convenience, we also define 
\begin{equation}
g_{e}\equiv\sqrt{g_{e_{L}}^{2}+g_{e_{R}}^{2}}\thinspace,\ g_{q}\equiv\sqrt{g_{q_{L}}^{2}+g_{q_{R}}^{2}}\thinspace,\label{eq:-52-1}
\end{equation}
and
\begin{equation}
r \equiv \frac{g_{\nu}}{g_{e}}\thinspace.\label{eq:r}
\end{equation}

Note that all effective couplings are of the order of $\epsilon_{{\rm loop}}g'$
except for the  mixing-induced part $SQ_{R}S^{\dagger}$ which is
of ${\cal O}(\sin^{2}\theta_{i})$ where the mixing angles $\theta_{i}$
are defined below Eq.~\eqref{eq:-21}. According to  Refs.~\cite{deGouvea:2015euy,Bolton:2019pcu},
for heavy neutrino masses above the electroweak scale, the current
constraints on $\sin^{2}\theta_{i}$ typically vary from $10^{-3}$
for $10^{-2}$, depending on lepton flavors. This allows for the scenario
that the mixing-induced part dominates over the loop-induced part
($\epsilon_{{\rm loop}}C_{L}C_{L}^{\dagger}$) in $g_{\nu}$.  More
concretely, if we consider $m_{R}\sim G_{F}^{-1/2}$, $\sin^{2}\theta_{i}\sim m_{D}^{2}/m_{R}^{2}$
should be of the same order as $G_{F}m_{D}^{2}$, which is about a
factor of $8\pi^{2}\sim{\cal O}(10^{2})$ larger than the loop-induced
part $\epsilon_{{\rm loop}}C_{L}C_{L}^{\dagger}$. In this case, one
expects that $g_{\nu}\sim{\cal O}(10^{2})g_{e}$. For larger $m_{R}$
with fixed $m_{D}$, the mixing-induced part decreases and eventually
looses its dominance, leading to $g_{\nu}\sim g_{e}$. Taking $m_{R}^{2}\in[1,\ 10^{2}]\thinspace G_{F}^{-1}$
and $\sin^{2}\theta_{i}=10^{-3}$ for example (neglecting the flavor
structure), we roughly have $m_{D}^{2}\in[10^{-3},\ 10^{-1}]\thinspace G_{F}^{-1}$
and $\epsilon_{{\rm loop}}\in[1.8\times 10^{-5},\ 1.8\times 10^{-3}].$
Therefore, as a benchmark in this work, we can consider 
\begin{equation}
C_{L}C_{L}^{\dagger}\rightarrow1\thinspace,\ SQ_{R}S^{\dagger}\rightarrow10^{-3}\thinspace,\ \epsilon_{{\rm loop}}\in[10^{-5},\ 10^{-3}]\thinspace.\label{eq:-5}
\end{equation}

\section{DUNE as a neutrino scattering experiment}\label{sec:DUNEnue}

Having formulated the effective couplings of $Z'$ to neutrinos and
electrons, we now study the sensitivity of the DUNE near detector
to the signal of elastic neutrino-electron scattering\footnote{Neutrino trident scattering could also be used to probe $Z'$. However,
	for our model the sensitivity is weaker than elastic neutrino-electron
	scattering, as can be expected from Fig.~7 of Ref.~\cite{Ballett:2019xoj}.} caused by $Z'$.

For a general $Z'$ with the effective couplings given by Eq.~\eqref{eq:-42}, the differential cross section of elastic   neutrino-electron
scattering including both the SM and the new physics contributions
reads~\cite{Lindner:2018kjo,Link:2019pbm,Marshall:2019vdy}:
\begin{equation}
\frac{d\sigma}{dT}=\frac{2m_{e}G_{F}^{2}}{\pi}\left[c_{L}^{2}+c_{R}^{2}\left(1-\frac{T}{E_{\nu}}\right)^{2}-c_{L}c_{R}\frac{m_{e}T}{E_{\nu}^{2}}\right],\label{eq:-43}
\end{equation}
where 
\begin{align}
c_{L} & =c_{L}^{({\rm SM})}+\frac{g_{eL}g_{\nu}}{2\sqrt{2}G_{F}\left(2m_{e}T+m_{Z'}^{2}\right)}\thinspace,\ \ c_{L}^{({\rm SM})}=-\frac{1}{2}+s_{W}^{2}+\delta_{\alpha e}\thinspace,\label{eq:-44}\\
c_{R} & =c_{R}^{({\rm SM})}+\frac{g_{eR}g_{\nu}}{2\sqrt{2}G_{F}\left(2m_{e}T+m_{Z'}^{2}\right)}\thinspace,\ \ c_{R}^{({\rm SM})}=s_{W}^{2}\thinspace.\label{eq:-45}
\end{align}
Here $T$ denotes the recoil energy of the
electron; $c_{L}$ and $c_{R}$ are dimensionless quantities consisting
of the SM and the new physics contributions, as given by Eqs.~\eqref{eq:-44}
and \eqref{eq:-45}. We have added $\delta_{\alpha e}$ (where $\alpha$
denotes the the incoming neutrino flavor) in Eq.~\eqref{eq:-44} to
take account of the SM charged current interaction: $\delta_{\alpha e}=1$
for $\alpha=e$ and $\delta_{\alpha e}=0$ for $\alpha=\mu$ or $\tau$.
Note that $c_{R}^{({\rm SM})}$ is not affected by the presence
of the charged current contribution. For anti-neutrino scattering
($\overline{\nu}_{\alpha}+e$), Eq.~\eqref{eq:-43} should be modified
with the interchange $c_{L}\leftrightarrow c_{R}$ while Eqs.~\eqref{eq:-44}
and \eqref{eq:-45} remain  the same. In this work, we neglect flavor
transition in neutrino scattering since the contribution is suppressed
due to the loss of interference with the SM processes.

The event rate of elastic neutrino-electron scattering at the detector
is computed by~\cite{Lindner:2018kjo,Link:2019pbm}:
\begin{equation}
\frac{dN}{dT}=N_{e}\lambda_{{\rm POT}}\int\Phi(E_{\nu})\frac{d\sigma(T,\ E_{\nu})}{dT}\Theta(T_{\max}-T)dE_{\nu}\thinspace,\label{eq:-47}
\end{equation}
where $N_{e}$ is the total electron number in the fiducial mass of the detector;
$\lambda_{{\rm POT}}$ is the number of protons-on-target (POT) at
the neutrino production facility; $\Phi$ is the neutrino flux per
POT; $\Theta$ is the Heaviside theta function with the maximal recoil energy
$T_{\max}$ determined by
\begin{equation}
T_{\max}(E_{\nu})=\frac{2E_{\nu}^{2}}{m_{e}+2E_{\nu}}\thinspace.\label{eq:-48}
\end{equation}
The event number in a given recoil energy bin $T\in[T_{i},T_{i}+\Delta T]$
is computed as 
\begin{equation}
N_{i}=\int_{T_{i}}^{T_{i}+\Delta T}\frac{dN}{dT}dT\thinspace.\label{eq:-49}
\end{equation}
The electron number $N_{e}$ can be computed as $N_{e}=18M_{{\rm Ar}}/m_{{\rm Ar}}$
where the factor $18$ comes from the fact that each ${\rm Ar}$ atom has 18 electrons,
$M_{{\rm Ar}}$ is the fiducial detector mass  and $m_{{\rm Ar}}=39.95u=37.21\ {\rm GeV}$ is the atomic mass of $^{40}$Ar.
For the fiducial mass, we consider only the liquid argon time-projection
chamber (LArTPC) near detector which has a fiducial mass of $M_{{\rm Ar}}=67.2$
ton \cite{DUNE:2020fgq}. For the number of POT, we take $\lambda_{{\rm POT}}=5\times1.1\times10^{21}$
for each of the neutrino and anti-neutrino modes, corresponding to
the Long Baseline Neutrino Facility (LBNF) at Fermilab operating at
1.2 MW intensity for five years in each mode. 
The neutrino flux $\Phi$
can be obtained from the latest DUNE Technical Design Report\footnote{See Fig.~5.4 in Ref.~\cite{DUNE:2020lwj}, Fig.~2.15 in Ref.~\cite{DUNE:2021tad}, and
	Fig.~2 in Ref.~\cite{DUNE:2021cuw}.}.


With the above setup, we compute the event numbers\footnote{Our code is publicly available at \url{https://github.com/xunjiexu/Zprime-at-DUNE}. }
and perform the sensitivity study with the following $\chi^{2}$ function
previously adopted in Ref.~\cite{Lindner:2016wff}:
\begin{equation}
\chi^{2}=\left(\frac{a}{\sigma_{a}}\right)^{2}+\sum_{i}\left(\frac{(1+a)N_{i}-N_{i}^{0}}{\sigma_{i}}\right)^{2},\label{eq:-50}
\end{equation}
where $a$ is a scale factor introduced to take into account the normalization
uncertainty (we assume that the dominant systematic uncertainty arises from the normalization of event rates); $\sigma_{a}$ denotes the uncertainty of $a$ for which we set $\sigma_a=2\%$; the summation
goes over all $T$ bins; $N_{i}$ denotes the event number evaluated
from Eq.~\eqref{eq:-49} with $N_{i}^{0}$ the corresponding SM value;
and $\sigma_{i}\approx\sqrt{N_{i}}$ denotes the statistical uncertainty on $N_{i}$.
Here $a$ is treated as a nuisance parameter which in the {\it frequentist
	treatment}~\cite{ParticleDataGroup:2020ssz} is marginalized by minimizing $\chi^{2}$ with
respect to $a$. For Eq.~\eqref{eq:-50}, the value of $a$ at the
minimum, $a_{\min}$, can be computed analytically \cite{Lindner:2016wff}:
\begin{equation}
a_{\min}=\frac{\sum_{i}(N_{i}^{0}-N_{i})N_{i}/\sigma_{i}^{2}}{\sigma_{a}^{-2}+\sum_{i}N_{i}^{2}/\sigma_{i}^{2}}\thinspace.\label{eq:-51}
\end{equation}
We set the $T$ bins as follows: the interval $T\in [0,\ 8]\ {\rm GeV}$ is divided evenly to 5 bins, and events above $8$ GeV are collected in a single bin. Each bin has a sufficiently large event number so that the use of $\sigma_{i}\approx\sqrt{N_{i}}$ is justified. The $\chi^2$ functions for neutrino and anti-neutrino modes are computed separately and then combined. 

\begin{figure}
	\centering

	\includegraphics[width=0.75\textwidth]{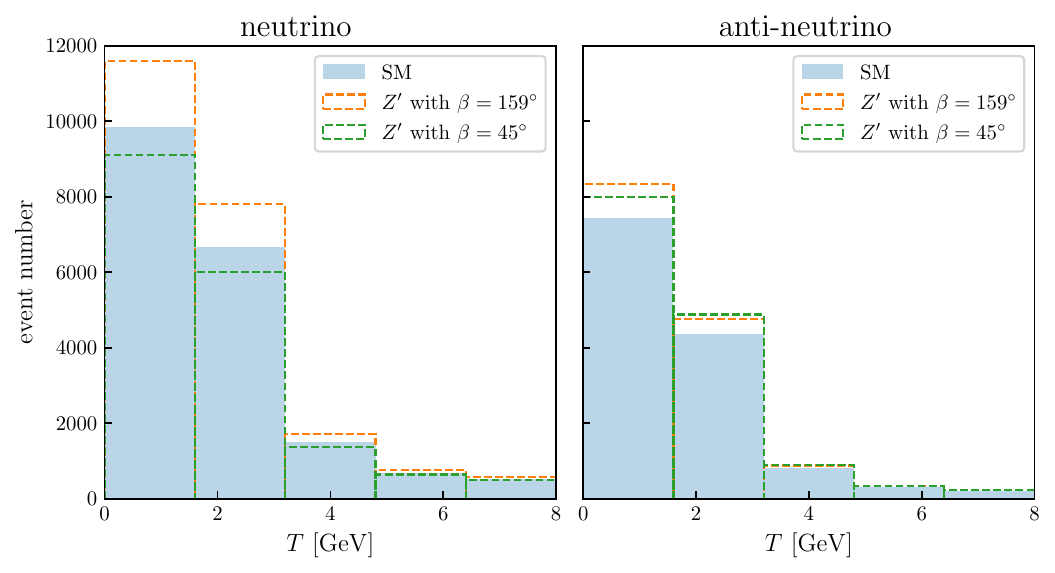} 
	\caption{ The events numbers for elastic neutrino-electron scattering at DUNE near detector LArTPC. Depending on the angle $\beta$ defined in Eq.~\eqref{eq:beta},  a generic $Z'$ can lead to an excess or deficit in the event numbers. 
		\label{fig:nu-e-events}} 
\end{figure}
\begin{figure}
	\centering

	\includegraphics[width=0.48\textwidth]{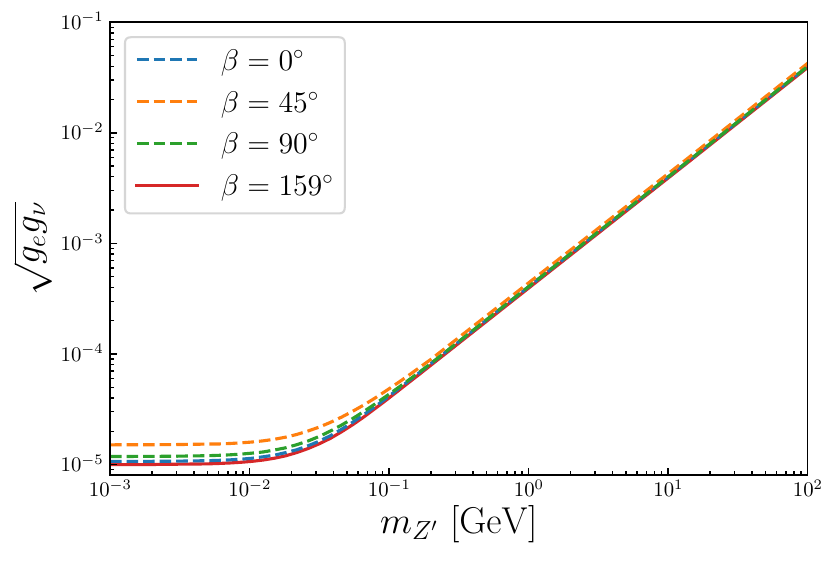} \includegraphics[width=0.48\textwidth]{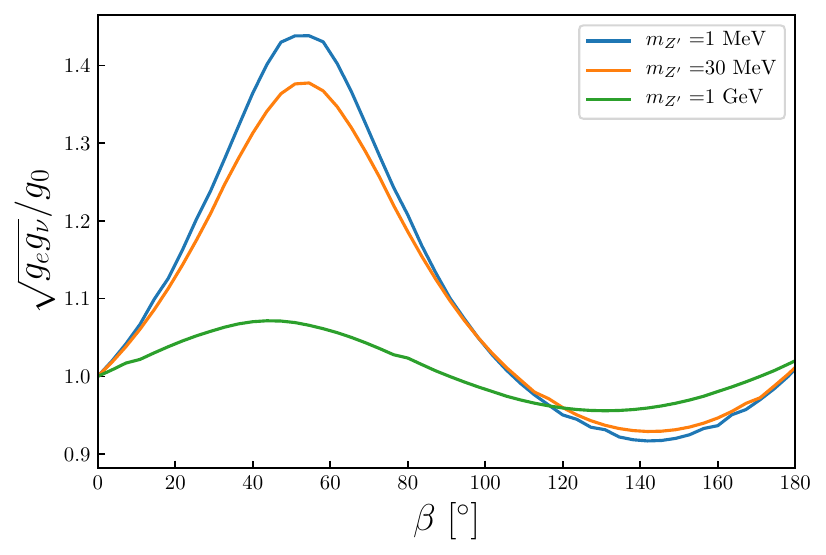}
	\caption{Left: the sensitivity of DUNE as a neutrino scattering experiment to a generic
		$Z'$ for several selected values of $\beta$ defined in Eq.~\eqref{eq:beta}. Right: the dependence of the results on $\beta$. 
		\label{fig:nu-e}}
\end{figure}


Fig.~\ref{fig:nu-e-events} shows the expected event numbers computed
in the SM and in the new physics scenario of a generic $Z'$. We parametrize the ratio of
$g_{e_{L}}$ to $g_{e_{R}}$ as 
\begin{equation}
(g_{e_{L}},\ g_{e_{R}})=(\cos\beta,\ \sin\beta)g_{e}\thinspace.\label{eq:beta}
\end{equation}
In our model, the couplings in Eq.~(\ref{eq:-53-2}) correspond
to $\beta=159^{\circ}$. This leads to excesses of event numbers in both 
neutrino and anti-neutrino modes, as shown by the orange points
in Fig.~\ref{fig:nu-e-events}. For comparison, we also show an example
with $\beta=45^{\circ}$ (corresponding to $g_{e_{L}}:g_{e_{R}}=1:1$)
which could cause excesses in the anti-neutrino mode but deficits
in the neutrino mode. The two new physics examples assume $m_{Z'}=100$
MeV and $\sqrt{g_{e}g_{\nu}}=10^{-4}$. 

In the left panel of Fig.~\ref{fig:nu-e}, we present the 90\% C.L.
sensitivity reach of DUNE-LArTPC for a few selected values of $\beta$.
As one can see, the curves nearly overlap for $m_{Z'}\gtrsim100$
MeV and show small differences for $m_{Z'}\lesssim10$ MeV. To further
check whether varying $\beta$ could cause significant changes, in
the right panel we investigate the dependence of the results on $\beta$
with three representative values of $m_{Z'}$  
and find that the sensitivity reach 
varies at most by a factor of $\sqrt{g_{e}g_{\nu}}/g_0 \in [0.91,1.44]$
for $\beta\in[0,\ \pi]$, where  $g_0$ is defined as the value of $\sqrt{g_{e}g_{\nu}}$ at $\beta=0$.
For larger $m_{Z'}$, the results are more  insensitive to $\beta$. 
Fig.~\ref{fig:nu-e} can be considered as a model-independent analysis
for $Z'$ with rather general couplings to left-/right-handed electrons and neutrinos.

\section{DUNE as a beam dump experiment}\label{sec:DUNEbd}

At the neutrino production site of DUNE, $Z'$ can be produced
from the proton beam striking the target.  Due to its weak couplings
to SM fermions, the produced $Z'$ boson can be long-lived and penetrate
through the shielding and earth between the near detector and the
target. This possibility allows us to consider DUNE as a beam dump
experiment (like E137, E141, Orsay) to probe $Z'$~\cite{Berryman:2019dme,Dev:2021qjj}.

We consider the DUNE Multi-Purpose Detector (MPD) which is a High-Pressure Ar gas TPC (HPgTPC)\footnote{Compared to LArTPC, HPgTPC has the advantage of lower (roughly 50 times smaller) backgrounds. Hence it is more suitable for new physics searches that do not require interactions with the content filled in the detector.} 
for the detection of $Z'$
in this work. 
The DUNE MPD will be located $579$ m away from the 
target, with a diameter of 5 m and a length of 5 m~\cite{Berryman:2019dme}.
Therefore, for $Z'$ particles produced at the target to cause observable
signals at the DUNE MPD, the transverse momentum, $p_{\perp}$, 
compared to the $Z'$ momentum, $p_{Z'}$, needs to be sufficiently
small:
\begin{equation}
\frac{p_{\perp}}{p_{Z'}}<\frac{R}{L_{1}}\approx4.3\times10^{-3}\equiv\theta_{{\rm max}}\thinspace,\label{eq:-8}
\end{equation}
where $R=2.5$ m is the radius of the detector and $L_{1}=579$ m
is the distance of the detector to the target. 

\begin{figure}
	\centering
	
	\includegraphics[width=0.75\textwidth]{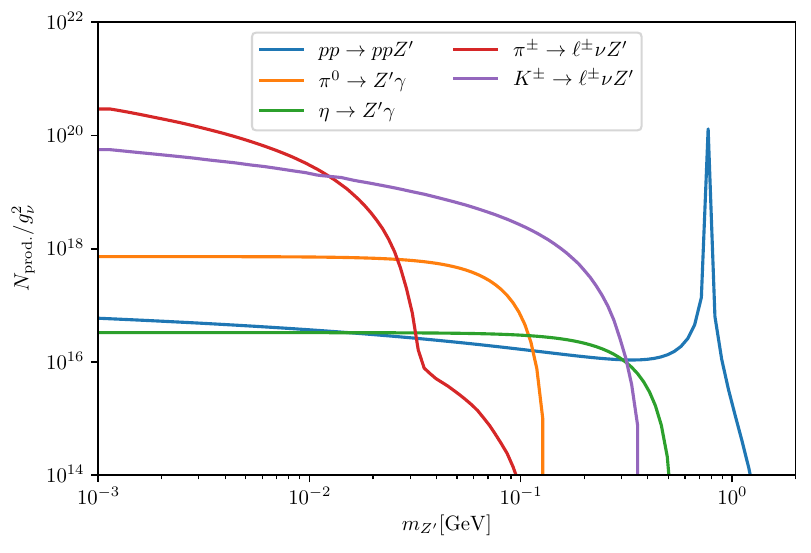}
	
	\caption{\label{fig:flux}The number of $Z'$ bosons produced via meson decays ($\pi^{0}\rightarrow Z'\gamma$,
		$\eta\rightarrow Z'\gamma$, $\pi^{\pm}\rightarrow\ell^{\pm}\nu Z'$,
		$K^{\pm}\rightarrow\ell^{\pm}\nu Z'$)  and proton bremsstrahlung
		($pp\rightarrow ppZ'$) assuming $r=100$ and
		$1.47\times10^{22}$ POT from a 120 GeV proton beam.}
\end{figure}

There are several processes responsible for the production of $Z'$
in our model. The dominant ones are meson decays ($\pi^{0}\rightarrow Z'\gamma$,
$\eta\rightarrow Z'\gamma$, $\pi^{\pm}\rightarrow\ell^{\pm}\nu Z'$,
$K^{\pm}\rightarrow\ell^{\pm}\nu Z'$) and proton bremsstrahlung ($pp\rightarrow ppZ'$).
The latter can be computed using an approximate formula in Ref.~\cite{Blumlein:2013cua}---see
Appendix~\ref{sec:proton-brem} for a brief review. The former 
requires the production rates of relevant mesons, which can be obtained
from Monte-Carlo simulations of high-energy protons scattering off
nucleons. We adopt the production rates from Ref.~\cite{Berryman:2019dme}
(neutral meson results from Fig.~4.1, charged meson results from
Fig.~5.2) and rescale them according to the effective couplings of
$Z'$  in our model.

In addition, $Z'$ could be produced via $Z'$-photon mixing, electron bremsstrahlung, and parton scattering ($q\overline{q}\to Z'$). 
In our model, the photon-$Z'$ mixing is absent at the one-loop level. They could be generated at 
the two-loop level, but the effect is expected to be highly suppressed. 
The production via electron bremsstrahlung is possible, where the
electron could be either from some final states of proton-nucleon
scattering or from the target (proton-electron scattering). The
former is subdominant compared to proton bremsstrahlung because electron
bremsstrahlung requires an electron in the final states while proton 
bremsstrahlung can appear in all kinds of proton scattering processes.\footnote{Here we would like to refer to Ref.~\cite{Celentano:2020vtu} which performed
Monte Carlo simulations to investigate the contribution of the secondary
production such as $e^{\pm}$ bremsstrahlung and annihilation. According
to Fig.~13 of Ref.~\cite{Celentano:2020vtu}, the secondary contribution is indeed subdominant
above a few MeV.} 
The latter is also subdominant for us because proton-nucleon scattering has
a larger cross section than proton-electron scattering. As for parton interactions like $q\overline{q}\to Z'$, it
is only significant at high energies above the GeV scale. As an example,
we would like to refer to a previous study for the SHiP experiment~\cite{SHiP:2020vbd}. As one can see from Fig.~13 of Ref.~\cite{SHiP:2020vbd} (the blue curve in
the upper panel), this process only causes a small correction to the high-mass tip. 
Since the proton beam of SHiP is much more energetic
than that of DUNE (400 GeV versus 120 GeV), we expect that the
contribution of this process should be less significant at DUNE.

Fig.~\ref{fig:flux} shows  the number of $Z'$ bosons produced (denoted
as $N_{{\rm prod.}}$)  via the aforementioned processes for $r=100$,
within the small angle determined by Eq.~\eqref{eq:-8}. Since in
this case $Z'$ is dominantly coupled to neutrinos, in the low-mass
regime the dominant production processes are those with neutrino final
states ($\pi^{\pm}\rightarrow\ell^{\pm}\nu Z'$, $K^{\pm}\rightarrow\ell^{\pm}\nu Z'$).
Note, however, that these  three-body decay processes are more
suppressed in the phase space than neutral meson decays ($\pi^{0}\rightarrow Z'\gamma$,
$\eta\rightarrow Z'\gamma$). Hence if all the couplings are of the
same order of magnitude, neutral meson decays would dominate over
charged meson decays~\cite{Dev:2021qjj}. When $m_{Z'}$ is greater
than the meson masses, proton bremsstrahlung starts to dominate. In
particular, the blue curve in Fig.~\ref{fig:flux} peaks at $m_{Z'}\approx0.8$
GeV due to the resonance caused by the $\rho$/$\omega$ meson~\cite{deNiverville:2016rqh,Faessler:2009tn}.

The number of detectable events, $N_{{\rm det.}}$, can be computed
by~\cite{Coy:2021wfs}
\begin{equation}
N_{{\rm det.}}=\int dp_{Z'}\frac{dN_{{\rm prod.}}(p_{Z'})}{dp_{Z'}}\thinspace P_{{\rm decay}}(p_{Z'})\thinspace{\rm BR}_{Z'\rightarrow{\rm vis.}}\thinspace,\label{eq:-33}
\end{equation}
where ${\rm BR}_{Z'\rightarrow{\rm vis.}}$ denotes the branching
ratio of $Z'$ decaying to visible states and $P_{{\rm decay}}$ denotes
the probability that a single $Z'$ particle travels from the target
to the detector and then decays in the detector. 
The momentum distribution of $Z'$ at the production, $dN_{{\rm prod.}}/dp_{Z'}$,
includes contributions from meson decay processes and proton bremsstrahlung.
The latter is elaborated in Appendix~\ref{sec:proton-brem} and the
former can be obtained by combining the kinematics of meson decays
at rest  and Lorentz boosts according to the meson momentum distributions
taken from Fig.~4 in Ref.~\cite{Kopp:2006ky}. 
The probability $P_{{\rm decay}}$ is computed by~\cite{Coy:2021wfs}
\begin{equation}
P_{{\rm decay}}=e^{-L_{1}/L_{Z'}}\left(1-e^{-L_{2}/L_{Z'}}\right),\label{eq:-34}
\end{equation}
with $L_{1}=579$ m, $L_{2}=5$ m, and $L_{Z'}$ denotes the distance
of flight before $Z'$ decay. The physical meaning of Eq.~\eqref{eq:-34}
is manifest: it is the probability of $Z'$ not decaying in $L_{1}$
multiplied by the probability of $Z'$ decaying in $L_{2}$. The distance
of flight, $L_{Z'}$, is computed by
\begin{equation}
L_{Z'}=\frac{\tau_{Z'}v}{\sqrt{1-v^{2}}}\thinspace,\label{eq:-36}
\end{equation}
where $1/\sqrt{1-v^{2}}$ is the Lorentz boost factor with 
the velocity $v=p_{Z'}/\sqrt{m_{Z'}^{2}+p_{Z'}^{2}}$. The lifetime
$\tau_{Z'}$ takes into account all decay modes:
\begin{equation}
\tau_{Z'}^{-1}=\Gamma_{Z'}=\Gamma_{Z'\rightarrow{\rm had.}}+\sum_{f\neq q}\Gamma_{Z'\rightarrow f\overline{f}}\thinspace,\label{eq:-37}
\end{equation}
where $\Gamma_{Z'\rightarrow{\rm had.}}$ denotes the decay width
of $Z'$ to hadronic states. The decay widths to other states $\Gamma_{Z'\rightarrow f\overline{f}}$
($f\neq q$) are computed analytically (see e.g. Ref.~\cite{Blumlein:2013cua}):
\begin{equation}
\Gamma_{Z'\rightarrow f\overline{f}}=\frac{g_{f}^{2}}{12\pi}m_{Z'}\left(1+2\frac{m_{f}^{2}}{m_{Z'}^{2}}\right)\sqrt{1-\frac{4m_{f}^{2}}{m_{Z'}^{2}}}\thinspace.\label{eq:-38}
\end{equation}
The hadronic decay width, $\Gamma_{Z'\rightarrow{\rm had.}}$, can
be computed using experimental measurements of the hadron-to-muon
cross section ratio in $e^{+}e^{-}$ collisions. For $\Gamma_{Z'\rightarrow{\rm had.}}$,
we adopt the same method previously used in Ref.~\cite{Coy:2021wfs}.  The visible decay width in Eq.~\eqref{eq:-33} takes the form
\begin{equation}
{\rm BR}_{Z'\rightarrow{\rm vis.}}\equiv 1-\frac{\Gamma_{Z'\rightarrow{\rm \nu \overline{\nu}}}}{\Gamma_{Z' }}\,.
\end{equation}
The effective coupling to neutrinos, $g_{\nu}$, plays a particularly
important role here because it reduces the visible decay width of $Z'$. When $g_{\nu}\gg g_e$ and $g_q$, we have $ {\rm BR}_{Z'\rightarrow{\rm vis.}} \ll 1$ and hence a significantly suppressed event rate.

\begin{figure}
	\centering
	
	\includegraphics[width=0.7\textwidth]{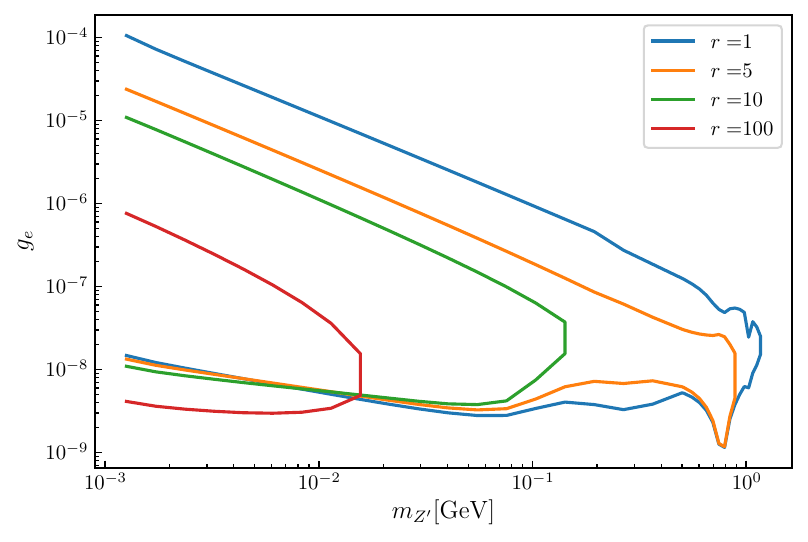}
	
	\caption{\label{fig:bird}The sensitivity reach of DUNE MPD to the $\nu_{R}$-philic
		$Z'$ with loop-induced couplings. The results depend on the ratio $r$  defined in Eq.~\eqref{eq:r}. }
\end{figure}

Using Eq.~\eqref{eq:-33} with each ingredient explicated above, we
compute the sensitivity reach of DUNE MPD for our model.  
The results are presented in Fig.~\ref{fig:bird} with  $r=g_{\nu}/g_e$ fixed at several benchmark values $(1,\ 5,\ 10,\ 100)$. As expected, when $r$ increases, due to the aforementioned suppression of the visible decay width, the  DUNE MPD sensitivity becomes weaker.

\section{Combined results and discussions}\label{sec:Results}

\begin{figure}
	\centering
	
	\includegraphics[width=0.49\textwidth]{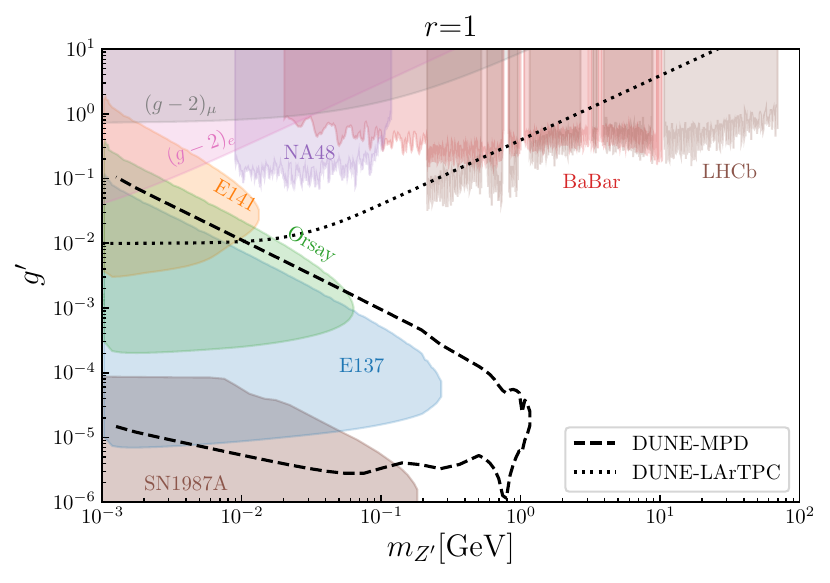}
	\includegraphics[width=0.49\textwidth]{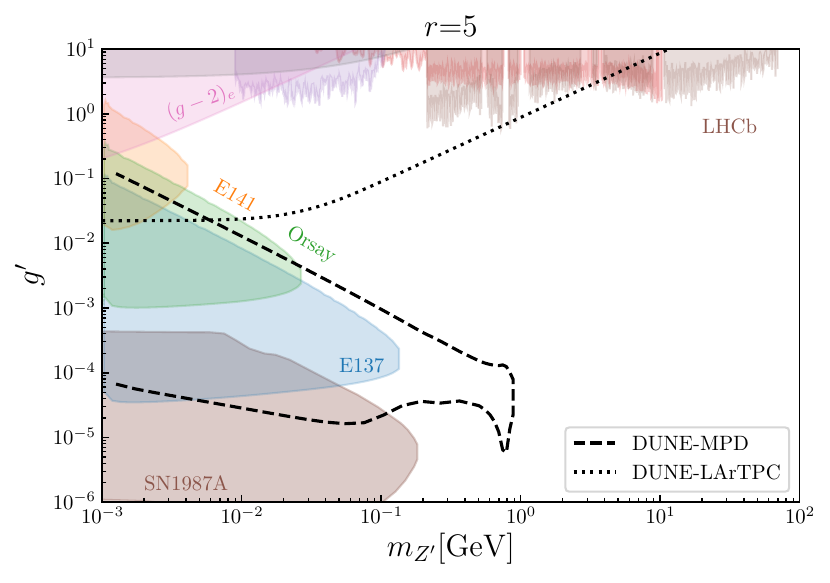}
	
	\includegraphics[width=0.49\textwidth]{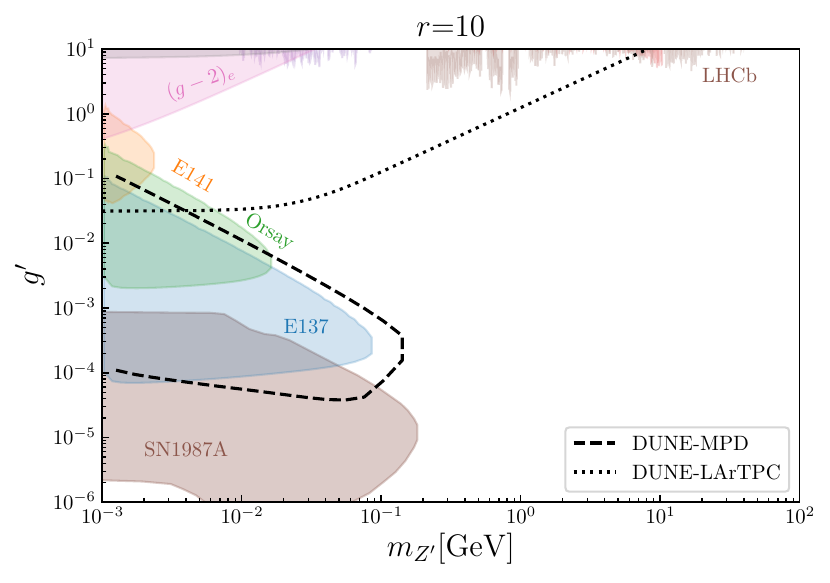}\includegraphics[width=0.49\textwidth]{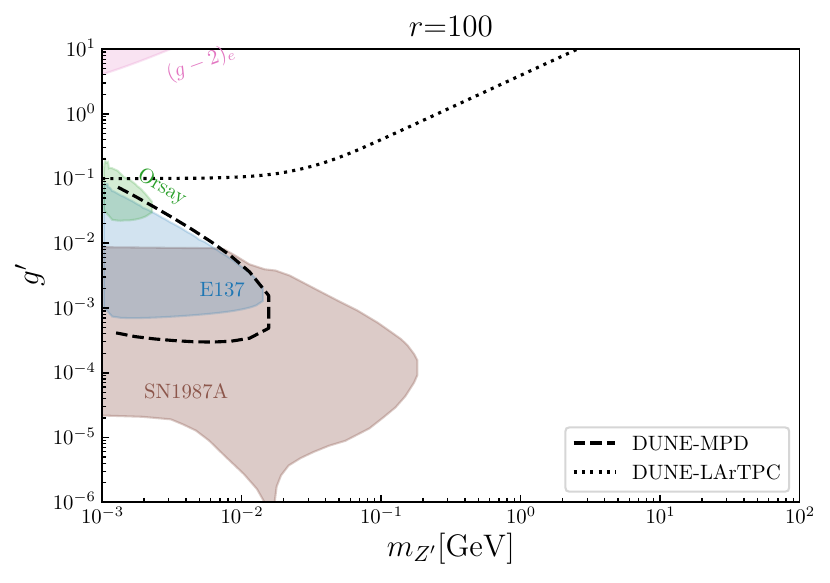}
	
	\caption{\label{fig:varying_r}Combined results of the DUNE sensitivity to
		the $\nu_{R}$-philic $Z'$. All bounds depends on the ratio $r\equiv g_{\nu}/g_{e}=\{1,\ 5,\ 10,\ 10^{2}\}$.}
\end{figure}

In this section, we compare the estimated sensitivities of DUNE-LArTPC (via $\nu$-$e$ scattering) and DUNE-MPD (which could detect $Z'$ decay) with existing bounds. 
We focus on the mass range $1\ \text{MeV}\leq m_{Z'}\leq100\ \text{GeV}$ which
covers the DUNE sensitivity reach.


For masses below 100 GeV, direct searches for $Z'$ at BaBar, LHCb, and NA48 put the strongest collider bounds. For masses ranging from MeV to GeV, there are constraints from beam dump experiments such as E137, E141, and Orsay.  For these existing collider and beam dump bounds, we use the {\tt DARKCAST} package~\cite{Ilten:2018crw} to recast the published bounds and obtain the shaded exclusion regions in Fig.~\ref{fig:varying_r}.

Since $Z'$ is effectively coupled to charged leptons, the muon and electron anomalous magnetic moments can be used to set constraints, which are presented in Fig.~\ref{fig:varying_r}, being labelled as   $(g-2)_\mu$ and $(g-2)_e$, respectively. These curves are obtained by setting $\Delta\alpha_{\mu}\leq5.5\times10^{-9}$ (5$\sigma$)~\cite{Muong-2:2021ojo}
and $\Delta\alpha_{e}\leq13.8\times10^{-13}$ (3$\sigma$)~\cite{Morel:2020dww}
where $\Delta\alpha_{\ell}$ ($\ell=\mu$, $e$) is computed by~\cite{Leveille:1977rc,Coy:2021wfs}
\begin{equation}
\Delta\alpha_{\ell}=\frac{g_{\ell}^{2}}{4\pi^{2}}\left(\frac{m_{\ell}}{m_{Z'}}\right)^{2}\int_{0}^{1}\frac{(1-x)x^{2}}{1-x+x^{2}(m_{\ell}/m_{Z'})^{2}}dx\thinspace.\label{eq:g-2}
\end{equation}

There are also strong constraints on low-mass $Z'$ from astrophysics and cosmology. The most relevant one comes from the supernova event SN1987A. Studies on SN1987A energy loss criteria have put restrictive constraints on weakly coupled $Z'$ below ${\cal O}(100)$ MeV~\cite{Dent:2012mx,Rrapaj:2015wgs,Chang:2016ntp}. 
In Fig.~\ref{fig:varying_r}, we adopt the combined SN1987A bound from Ref.~\cite{Knapen:2017xzo} 
for the $B-L$ gauge boson  and rescale it accordingly. Near the MeV mass scale or lower, there are constraints from stellar cooling ~\cite{Redondo:2013lna} 
and the effective number of relativistic degrees of freedom ($N_{\rm eff}$)~\cite{Escudero:2019gzq}.  We do not show the $N_{\rm eff}$ constraints as they are marginally relevant to the mass range considered here.

From Fig.~\ref{fig:varying_r}, we see that the ratio $r\equiv g_{\nu}/g_{e}$  is crucial when comparing bounds from DUNE and the existing experiments. 
When $g_{\nu}$ and $g_e$ are comparable ($r=1$), neutrino-electron scattering at DUNE-LArTPC (dotted line) does not show a significant advantage over other experiments. Nevertheless, there is a mass range from 10 MeV to 200 MeV for DUNE-LArTPC to probe. When $r$ increases, both the DUNE-LArTPC sensitivity and collider bounds become weaker, since 
they all depend on $g_e$ and/or $g_q$ which is decreasing, given fixed mixing angles and hence fixed $g_{\nu}/g'$. 
However, as one can see, for large $r$,  DUNE-LArTPC shows significantly better sensitivity  than other non-neutrino experiments because the former relies on the product $g_e g_{\nu}$ while the latter typically relies on $g_e^2$ or $g_q^2$. We emphasize here that the main feature of our framework is the dominance of neutrino couplings in the $\nu_R$-philic $Z'$ model, which provides an excellent prospect for DUNE to probe.

As previously discussed in Sec.~\ref{sec:DUNEbd}, DUNE-MPD can be used to probe $Z'$ like a beam dump experiment. The sensitivity reach is shown as dashed curves in Fig.~\ref{fig:varying_r}. In this case, $g_{\nu}$ plays a positive role in production (it leads to additional contributions of $Z'$ production from meson decays with neutrino final states), but a negative role in detection because it increases the invisible decay width of $Z'$. In particular, for $r\gg 1$, $Z'$ dominantly decays to neutrinos. Overall, large $g_{\nu}$ reduces the sensitivity of DUNE-MPD. Therefore, when $r$ increases, 
DUNE-MPD together with all other beam dump experiments quickly looses sensitivity to the weak-coupling regime ($g'\sim 10^{-4},\ 10^{-5}$). Consequently,  the DUNE-MPD sensitivity region shrinks quickly as we increase $r$ in Fig.~\ref{fig:varying_r}.

In summary, DUNE-LArTPC can probe the $\nu_{R}$-philic $Z'$ with
a mass up to $\sim10$ GeV when the coupling $g'$ approaches the
perturbativity bound, while DUNE-MPD can probe it up to $\sim1$ GeV,
mainly limited by the beam energy. For smaller masses, DUNE-LArTPC
can probe $g'$ down to $\sim10^{-2}$ while DUNE-MPD can probe in
general much weaker couplings, possibly reaching $\sim10^{-6}$ for
certain masses.





\section{Conclusion\label{sec:Conclusion}}

Hidden $U(1)$ symmetries in the right-handed neutrino ($\nu_R$) sector are theoretically well motivated and would give rise to an inherently dark gauge boson which we refer to as the $\nu_R$-philic $Z'$. 
Due to the loop-suppressed couplings to normal matter and comparatively larger couplings to neutrinos, neutrino experiments are the most suited to probe the $\nu_R$-philic $Z'$. 

In this work we studied the sensitivity of the future DUNE experiment to the $\nu_R$-philic $Z'$. 
We considered two complementary near detectors,  DUNE-LArTPC and DUNE-MPD (HPgTPC), which could be sensitive to  $Z'$ signals via elastic $\nu$-$e$ scattering and via $Z'$ decay respectively. We stress here that the ratio of electron and neutrino couplings, $r=g_\nu/g_e$, which is practically a free parameter in the model, plays a crucial role. Larger neutrino couplings lead to higher elastic $\nu$-$e$ scattering rates in DUNE-LArTPC but make $Z'$ decay less visible in  DUNE-MPD due to the enhanced invisible decay width. The combined results are shown in Fig.~\ref{fig:varying_r} and compared with existing bounds. For $r=1$ or $5$, DUNE-MPD exhibits a significant advantage over other beam dump experiments in the mass range $0.1 {\rm GeV} \lesssim m_{Z'}\lesssim 1 {\rm GeV}$. On the other hand, for larger $r$ such as $r=10$ or $100$, DUNE-LArTPC as a scattering experiment will be able to generate the leading constraints, exceeding the collider bounds from BaBar, LHCb, etc. We conclude that there is an excellent prospect of DUNE probing new physics hidden in the sector of right-handed neutrinos.


\begin{acknowledgments}
	We thank Vedran Brdar, Matheus Hostert, Kevin Kelly and Doojin Kim  for helpful
	discussions and assistance. The work of B.D. is supported in part by the U.S. Department of Energy under grant No.~DE-SC0017987 and by a Fermilab Intensity Frontier Fellowship. B.D. acknowledges the local hospitality at KITP where part of this work was done. The research at KITP was supported in part by the National Science Foundation under Grant No. NSF PHY-1748958. X.J.X is supported in part by the National Natural Science Foundation of China under grant No. 12141501.
\end{acknowledgments}

\appendix

\section{ The proton bremsstrahlung formula\label{sec:proton-brem}}

For $Z'$ emitted via proton bremsstrahlung ($pN\rightarrow pNZ'$),
there is an approximate formula often used in the literature~\cite{Blumlein:2013cua,deNiverville:2016rqh,Berryman:2019dme,SHiP:2020vbd}\footnote{We note that there is a discrepancy between the expressions of $w_{ba}$
	in Refs.~\cite{Blumlein:2013cua,SHiP:2020vbd} and Refs.~\cite{deNiverville:2016rqh,Berryman:2019dme}.
	We adopt the former since  this is consistent with the matrix element
	given by  Eq.~(6) in Ref.~\cite{Blumlein:2013cua}. }:
\begin{equation}
\frac{d^{2}N_{{\rm brem}}}{dzdp_{\perp}^{2}}=N_{{\rm POT}}\frac{\sigma_{pN}(s')}{\sigma_{pN}(s)}\left|F_{1,N}(m_{Z'}^{2})\right|^{2}w_{ba}(z,\thinspace p_{\perp}^{2})\thinspace,\label{eq:-9}
\end{equation}
with
\begin{align}
w_{ba}(z,\thinspace p_{\perp}^{2}) & =\frac{g_{p}^{2}}{8\pi^{2}H}\left[\frac{1+(1-z)^{2}}{z}-2z(1-z)\left(\frac{2m_{p}^{2}+m_{Z'}^{2}}{H}-z^{2}\frac{2m_{p}^{4}}{H^{2}}\right)\right.\nonumber \\
& \left.\quad +2z(1-z)\left[1+(1-z)^{2}\right]\frac{m_{Z'}^{2}m_{p}^{2}}{H^{2}}+2z(1-z)^{2}\frac{m_{Z'}^{4}}{H^{2}}\right],\label{eq:-10}\\
H & \equiv p_{\perp}^{2}+(1-z)m_{Z'}^{2}+z^{2}m_{p}^{2}\thinspace.\label{eq:-11}
\end{align}
Here the kinematic variable
$z$ is the ratio of the outgoing $Z'$ momentum to the incoming proton
momentum ($0<z<1$); $p_{\perp}$ is the transverse component of the
$Z'$ momentum;  $\sigma_{pN}$ is the proton-nucleus cross section
evaluated at $s=2m_{p}E_{p}$ and $s'\approx2m_{p}(E_{p}-E_{Z'})$;
$F_{1,N}(m_{Z'}^{2})$ is the vector form factor of the nucleus evaluated
at $q^{2}=m_{Z'}^{2}$ (timelike); and $g_{p}$ is the effective coupling
of $Z'$ to the proton. Since the beam energy is high, one can approximately
view the target as free protons and neutrons at rest. According to
the simulations in Ref.~\cite{Berryman:2019dme}, the difference between
neutron and proton being the target particle is negligible at this
energy,  which allows us to replace $\sigma_{pN}$ with the proton-proton
cross section, $\sigma_{pp}$, and replace $F_{1,N}$ with the corresponding
form factor of the proton, $F_{1,p}$. The cross section $\sigma_{pp}$
is computed via an approximate formula from Ref.~\cite{Blumlein:2013cua},
and the form factor $F_{1,p}$ is taken from Ref.~\cite{deNiverville:2016rqh}. 

Due to the small angle in Eq.~\eqref{eq:-8} which implies $p_{\perp}^{2}\ll H$,
we can integrate out $p_{\perp}^{2}$ in Eq.~\eqref{eq:-9}, which
leads to
\begin{equation}
\frac{dN_{{\rm brem}}}{dz}=N_{{\rm POT}}\frac{\sigma_{pN}(s')}{\sigma_{pN}(s)}\left|F_{1,N}(m_{Z'}^{2})\right|^{2}w_{ba}(z)\thinspace,\label{eq:-13}
\end{equation}
where
\begin{equation}
w_{ba}(z)=\int_{0}^{E_{p}^{2}\theta_{{\rm max}}^{2}}w_{ba}(z,\thinspace p_{\perp}^{2})dp_{\perp}^{2}=\frac{g_{p}^{2}\theta_{{\rm max}}^{2}E_{p}^{2}}{8\pi^{2}}\frac{z\left(z^{2}-2z+2\right)}{\left[m_{p}^{2}z^{2}+m_{Z'}^{2}(1-z)\right]}\thinspace.\label{eq:-12}
\end{equation}
Note that, though $w_{ba}$ in Eq.~\eqref{eq:-10} diverges at $z\rightarrow0$,
the integrated form of $w_{ba}$ in Eq.~\eqref{eq:-12} at $z\rightarrow0$
has been regularized by $m_{Z'}^{2}$, i.e. $\lim_{z\rightarrow0}w_{ba}(z)=0$
as long as $m_{Z'}^{2}\neq0$. 

The proton bremsstrahlung formula  presented above is valid only
if the incoming and outgoing protons as well as the $Z'$ boson are
ultra-relativistic~\cite{Blumlein:2013cua}. Typically in the literature
one imposes upper and lower bounds on $z$, $z_{\min}<z<z_{\max}$.
For instance, previous studies on the SHiP experiment often take $(z_{\min},\ z_{\max})=(0.1,\ 0.9)$~\cite{deNiverville:2016rqh,SHiP:2020vbd}.
For the DUNE setup, we set  $z_{\min}=5m_{Z'}/E_{p}$ so that the
outgoing $Z'$ is relativistic. If $z$ is very close to $1$, the
outgoing proton would be non-relativistic. Hence we set the upper
bound $z_{\max}=1-5m_{p}/E_{p}$. The factor $5$ here  is chosen
so that the relativistic approximation holds and meanwhile the width
of the integration  interval ($z_{{\rm max}}-z_{\min}\approx1$) is
not significantly reduced. We have checked that varying this factor
does not change the result significantly, provided that the two conditions
are satisfied. 

\bibliographystyle{JHEP}
\bibliography{ref}

\end{document}